\documentclass[aps,preprint,showpacs]{revtex4}
\usepackage{amsmath}
\usepackage{bm}
\usepackage{graphics}
\usepackage{graphicx}
\def  \d {DNA}
\def  \eps {\epsilon}
\begin{document}
\title{Statistical Theory of Force Induced Unzipping of DNA}
\author{Navin Singh and Yashwant Singh}
\affiliation{Department of Physics, Banaras Hindu University, \\
Varanasi - 221005, INDIA} 
\email{knavins@gmail.com}
\begin{abstract}
The unzipping transition under the influence of external force of a
ds\d\ molecule has been studied using the Peyrard-Bishop Hamiltonian.
The critical force $F_c(T)$ for unzipping calculated in the constant
force ensemble is found to depend on the potential parameter $k$ which
measures the stiffness associated with a single strand of \d\ and on
$D$, the well depth of the on-site potential representing the strength
of hydrogen bonds in a base pair. The dependence on temperature of 
$F_c(T)$ is found to be $(T_D - T)^{1/2}$ ($T_D$ being the thermal
denaturation temperature) with $F_c(T_D) = 0$ and $F_c(0) = \sqrt{2kD}$.
We used the constant extension ensemble to calculate the average force 
$F(y)$ required to stretch a base pair $y$ distance apart. The value of
$F(y)$ needed to stretch a base pair located far away from the ends of a
ds\d\ molecule is found twice the value of the force needed to stretch
a base pair located at one of the ends to the same distance for $y \geq 1.0 \; {\rm \AA}$.
The force $F(y)$ in both cases is found to have a very large value for 
$y \approx 0.2 \; {\rm \AA}$ compared to the critical force found from  the
constant force ensemble to which $F(y)$ approaches for large values of $y$. 
It is shown that the value of $F(y)$ at the peak depends on the value of $k\rho$ 
which measures the energy barrier associated with the reduction in \d\ strand 
rigidity as one passes from ds\d\ to ssDNA and on the value of the depth of the
on-site potential.
The effect of defects on the position and height of the peak in $F(y)$
curve is investigated by replacing some of the base pairs including the one being
stretched by defect base pairs.  The formation and behaviour
of a loop of Y shape when one of the ends base pair is stretched
and a bubble of ss\d\ with the shape of ``an eye'' when a base pair far from
ends is stretched are investigated.
\end{abstract}
\maketitle

\section{Introduction}
\label{sec:1}
Natural DNA is a giant double stranded linear molecule with length ranging
from 2$\mu m$ for simple viruses to $3.5 \times 10^7 \mu m$ for more complex
organism and is known to have a complex nature of internal motions \cite{yak}.
The structural elements such as individual atoms, group of atoms (bases, sugar
rings, phosphates) fragments of double chain including several base pairs, are
in constant movements and this movement plays crucial role in its biological
activities. During transcription a transient "bubble" of single stranded \d\
is formed, to allow enzymes that make a mRNA copy of \d\ sequence to access
the \d\ bases \cite{watson}. In a replication the separation of two strands
starts from one end and propagates to the other end; separated parts of each
strand serves as a template for the synthesis of a new strand and thus making
two exact copies of the \d . The energy involved in these processes is of the
order of 5-25 kcal/mole. These motions {\it in vitro} can be activated by
increasing temperature, changing pH and/or solvent conditions. The process of
separating the two strands wound in a double helix into two single strands
upon heating is known as thermal denaturation. Several experiments on dilute
\d\ solutions \cite{wart} have provided evidence for the existence of a
thermally driven melting transition corresponding to the sudden opening of
base pairs at a denaturation or melting temperature $T_D$.

In the living organisms the \d\ strands are forced open by proteins which
pull the strands of \d\ on selected positions. The recently developed
experimental techniques of micromanipulations and nanomanipulations
\cite{black,kish,hans} have now made it possible to probe the force
elongation characteristics of double stranded \d\ molecule (ds\d ),
determining its response to external force and torque {\it in vitro} at
temperatures where ds\d\ is thermally stable in absence of the external
force. The mechanical unzipping of ds\d\ by a force pulling the end of one of
the two strands, the end of the other strand being anchored to some physical
support (see Fig. 1) has been studied by Bockelmann and co-workers
\cite{roulet} who measured the average force for the opening of the two
strands. It has been shown that the two strands of a ds\d\ can be pulled apart
if a force $\approx 12$ pN  is applied with some variation  about this mean
value depending upon the sequence.
\begin{figure}[h]
\resizebox{0.4\textwidth}{!}{\includegraphics{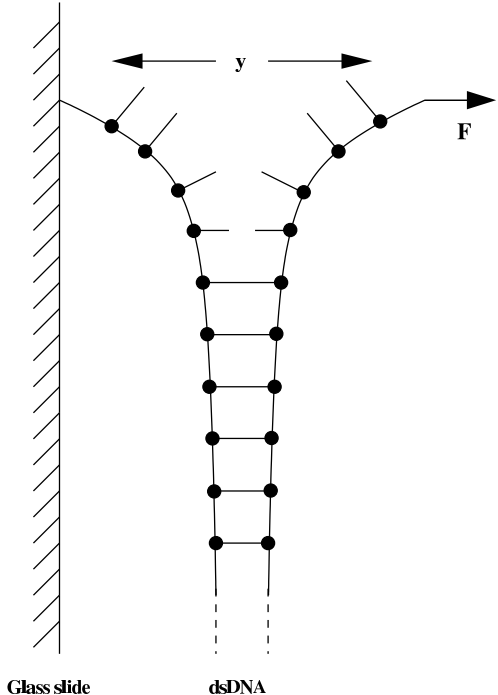} }
\caption{The schematic of DNA unzipping by force $F$ applied at one end of one of the two 
strands of a ds\d\ molecule and the end of the other strand attached to a stationary
glass plate.}
\label{fig:1}
\end{figure}
In single molecule experiments, the results may depend on the
statistical ensemble one works with \cite{somen1,luben}. It may be noted that
techniques like atomic force microscope \cite{ans} use the fixed extension
ensemble while the magnetic bead method \cite{danil} uses the fixed force
ensemble. In view of this we have studied the unzipping transitions in both
the fixed force (sec. 4) and the fixed extension (sec. 5) ensembles.
 
In order to make a theoretical approach feasible one has to reduce the
complexity of internal motions of ds\d\ to the minimum. Clearly, an
appropriate choice of the relevant degrees of freedom depends on the problem
one is interested. For example, the models based on the theory of polymers that
use self-avoiding walks to describe the two strands \cite{kafri} can
be very successful in studying the properties of melting transition at the
large scale but they cannot be used to investigate the properties that depend
on the sequence, or probe the \d\ at the microscopic scale as is done in some
single molecule experiments. One of the simplest models that investigate \d\
at the scale of base pairs is the Peyrard-Bishop model \cite{pb89,daux93}.
Though as described in Sec. 2, this model ignores the helicoidal structure
\cite{simona} of \d , it is found to have enough details to analyze
mechanical behaviour at the few ${\rm \AA}$ scale relevant to 
molecular-biological events.

A number of attempts have recently been made to understand various aspects of dsDNA
unzipping \cite{somen1,luben,thomp,somen2,cocco1,cocco2,somen3}.
Our aim in this work is to give
further insight on the various intricacies involved in the unzipping of
ds\d . In particular, we investigate how the unzipping transition depends on
the different parameters that appear in the model Hamiltonian and on the presence
of defect base pairs. The paper is
organized as follows: In Sec. 2 we describe in detail the different parts of
the Peyrard-Bishop (PB) model and the set of potential parameters that we use
in our study. In Sec. 3 we provide a brief outline of the theory that is
used to investigate the transitions in a homogeneous ds\d . This theory is 
used in Sec.  IV to calculate the value of critical force, $F_c(T)$, for 
unzipping as a function of temperature. The curve $F_c(T)$ gives the boundary 
separating
the zipped state from the unzipped state. The critical force is shown to
depend most predominantly on the stiffness associated with a single strand. In
Sec. 5 we investigate the unzipping of ds\d\ in a constant extension ensemble
and calculate the average force needed to maintain the extension.
It is shown that a very large force is needed to enforce an extension of
about $1 {\rm \AA}$ and the value of this force depends on the barrier
associated with reduction in \d\ strand rigidity as one passes from ds\d\ to
ss\d . The change in height and position of the barrier is investigated by
introducing defect base pairs. The extension can take place at any point along
the strand. We have discussed the two cases; the extension at one end and at
the middle. We summarize our results in Sec. 6.

\section{Model Hamiltonian and Potential Parameters}
\label{sec:2}

Since the internal motion that is predominantly responsible for unzipping of
ds\d\ in situation shown in Figure 1 is the stretching of the bases from their
equilibrium position along the direction of the hydrogen bonds that connect
the two bases, a \d\ molecule can be considered as a quasi one dimensional
lattice composed of $N$ base pair units. The forces which stabilize the
structure are the hydrogen bonds between complementary bases on opposite
strands and stacking interactions between nearest neighbour bases on opposite
strands. Each base pair is in one of the two states: either open(non hydrogen
bonded) or zipped (hydrogen bonded). In the presence of a force acting on one
end of the base pair, the Hamiltonian can be written as,
\begin{equation} 
H = \sum_{n=1}^N
H(y_n, y_{n+1}) - F\cdot y
\end{equation}
where $y_n$ denotes the stretching from the equilibrium position of the
hydrogen bonds connecting the two bases of the $n^{th}$ pair. A model
Hamiltonian that contains the minimum complexity of the internal motion at
base pairs level and accounts for the stability of ds\d\ structure is
\cite{pb89}.
\begin{equation} 
H(y_n, y_{n+1}) = \frac{p_n^2}{2m} + w(y_n, y_{n+1}) + V(y_n)
\end{equation}
where $m$ is the reduced mass of a base pair, and
\begin{equation}
p_n = m\left(\frac{dy_n}{dt}\right)
\end{equation}
The on-site potential $V(y_n)$ describes the interaction of the two bases of
the $n^{th}$ pair. The Morse potential
\begin{equation}
V(y_n) = D_n(e^{-ay_n} - 1)^2
\end{equation}
is generally taken to represent the on-site interaction. It may be
noted that $V(y_n)$ does not represent only the hydrogen bonds connecting
two bases belonging to opposite strands, but also the surrounding solvent
effects and the repulsion interactions of the phosphates. In Eq. (4)
$D_n$ measures the depth of the potential and $a$ its range. In a
homogeneous \d , $D_n$ is taken to be site independent but in a
heterogeneous (or natural) \d\ the value of $D_n$ depends on whether  the
$n^{th}$ base pair is AT or CG. The flat part at large values of the
displacement of this potential emulates the tendency of the pair "melt" at
high temperatures as thermal phonons drive the molecule outside the well and
towards the flat portion of the potential.

The stacking interactions are contributed by dipole-dipole interactions,
$\pi$-electron systems, London dispersion forces and in water solution, the
solvent induced hydrophobic interactions. These forces result in a complex
interaction pattern between overlapping base pairs, with minimum energy
distance close to 3.4 ${\rm \AA}$ in the normal \d\ double helix. As
suggested by Peyrard and Bishop \cite{daux93} the following anharmonic
potential model mimic the essential features of the stacking energy:
\begin{equation}
w(y_n, y_{n+1}) = \frac{1}{2} k \left[ 1 + \rho e^{-b(y_n+y_{n+1})}\right]
(y_n-y_{n+1})^2
\end{equation}
where the force constant $k$ is related with the stiffness of a single
strand and the second term in the bracket represents the anharmonic term.
The "anharmonic range" is defined by the parameter $b$. In the zipped
state the force constant is equal to $k(1+\rho)$. Decrease in the force
constant in the unzipped state provides a large entropy and hence favours
unzipping either at high force or at high temperatures. The difference in
the force constants between the zipped and the unzipped state of base
pairs, creates an energy barrier the value of which depends on $\rho$ and
range $b$.

The Hamiltonian model described above has five parameters $D_n$, $k$, $\rho$,
$a$ and $b$. In our calculation described below we have taken $a
= 4.2 \; {\rm \AA}^{-1} $ and $b = 0.35 \; {\rm \AA}^{-1}$.  For other three
parameters two different sets of values have been used; (i) $D_n = 0.11$ eV,
$\rho = 475.0 $, $k = 0.0032 $ eV${\rm \AA^{-2}}$ and
(ii)$D_n = 0.063$ eV, $\rho = 5.0 $, $k = 0.025 $ eV${\rm \AA}^{-2}$.
We shall henceforth refer them as potential parameters of set (i) or (ii).
The values of set (i) are close to those taken by Cocco et al \cite{cocco2} to
study the unzipping of \d\ at $T = 298$ K under the influence of force,
whereas those of set (ii) are close to those used by us \cite{nav} and by
others \cite{daux93,campa,theo} in the study of thermal denaturation of \d .
For the reduced mass $m$, a value of 300 a.m.u. has been used. For both sets
of parameters the denaturation temperature in absence of force is found close
to 350 K (see below) which, in turn is close to the value found for a natural
\d . The denaturation transition is found to be first order for both set of
parameters with finite melting entropy, a discontinuity in the fraction of
bound base pairs and divergent correlation lengths.

\section{Statistical theory for Unzipping transition}
\label{sec:3}
We consider a ds\d\ having $N$ number of base pairs at temperature $T$.
The classical partition function of it can be written as
\begin{equation}
Z_N(\beta) = Z_N^c(\beta) Z_N^k\left(\frac{\beta k}{2\pi}\right)^{-N/2}
\end{equation}
where
\begin{equation}
Z_N^k = (2\pi m k_BT)^{N/2}
\end{equation}
is the kinetic part. The configurational part of the partition function is
written as (taking $y_{N}$ to be fixed at zero value)
\begin{eqnarray}
Z_N^c  &=& \left(\frac{\beta k}{2\pi}\right)^{N/2} \int \left( \prod_{n=1}^N dy_n\right)
\delta(y_{N} - 0)  \nonumber \\
&\times & \exp \{-\beta\sum_{n=1}^{N-1} [\frac{1}{2}(V(y_n) + V(y_{n+1}) )
+ w(y_n,y_{n+1})  \nonumber  \\
& - &  F(y_n - y_{n+1}) ] \}
\end{eqnarray}
where $F$ is the force. Note that the factor $(\beta k/2\pi)^{-N/2}$ included in
Eq. (6) is balanced by its inverse included in Eq. (8).  For $N\to
\infty$ and force, $F$, acting on $n=1$ base pair, taking $y_{N} = 0$ is justified.
In denaturation study of \d\ when $F=0$ and the ends are free, the periodic
boundary condition, which amounts to adding a fictitious base pair with index $N+1$ 
having the same dynamics as base pair 1 (i.e. $y_{N+1} = y_1$), is imposed.

\begin{figure}
\includegraphics[height=3.25in,width=3.25in]{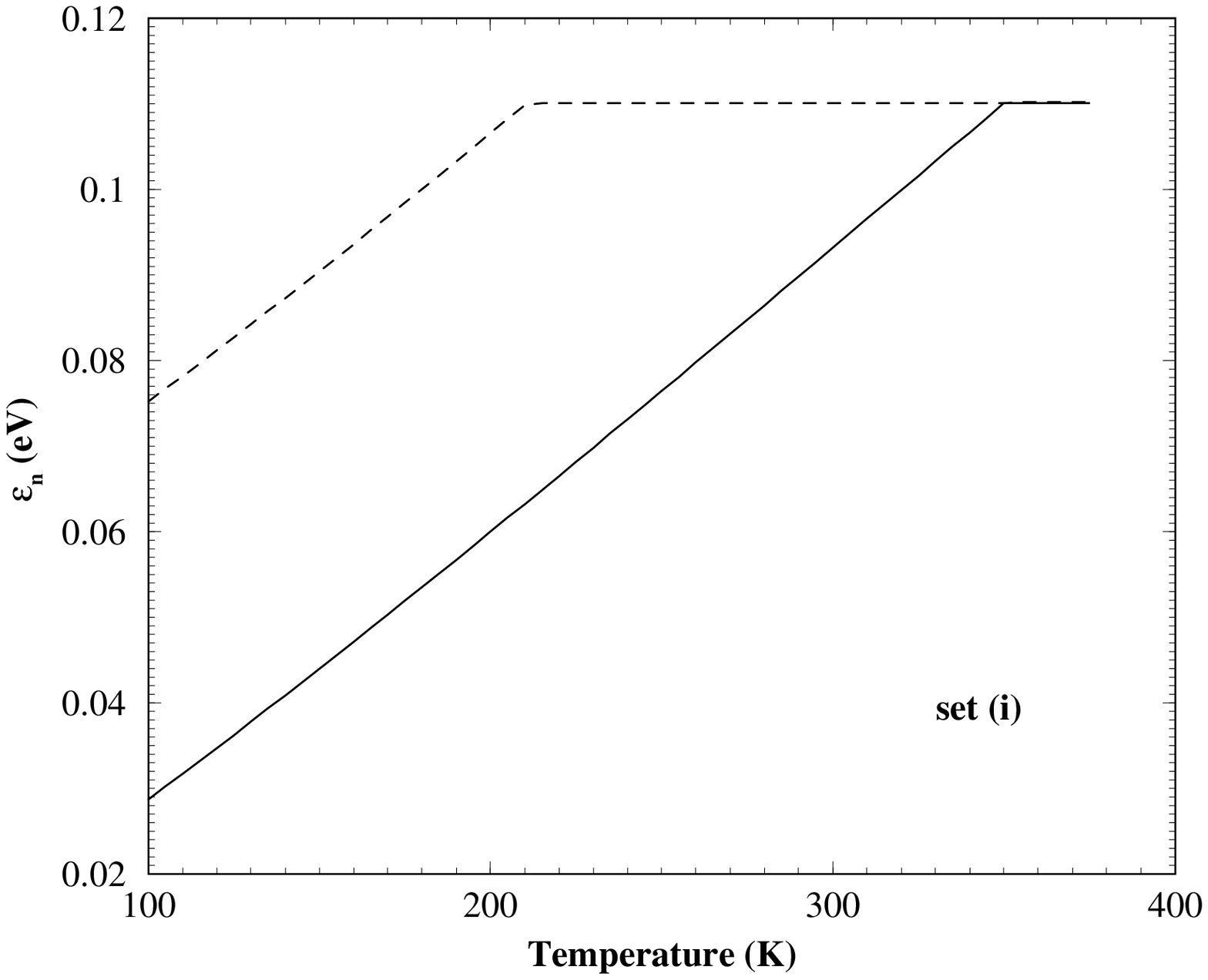} \\
\includegraphics[height=3.25in,width=3.25in]{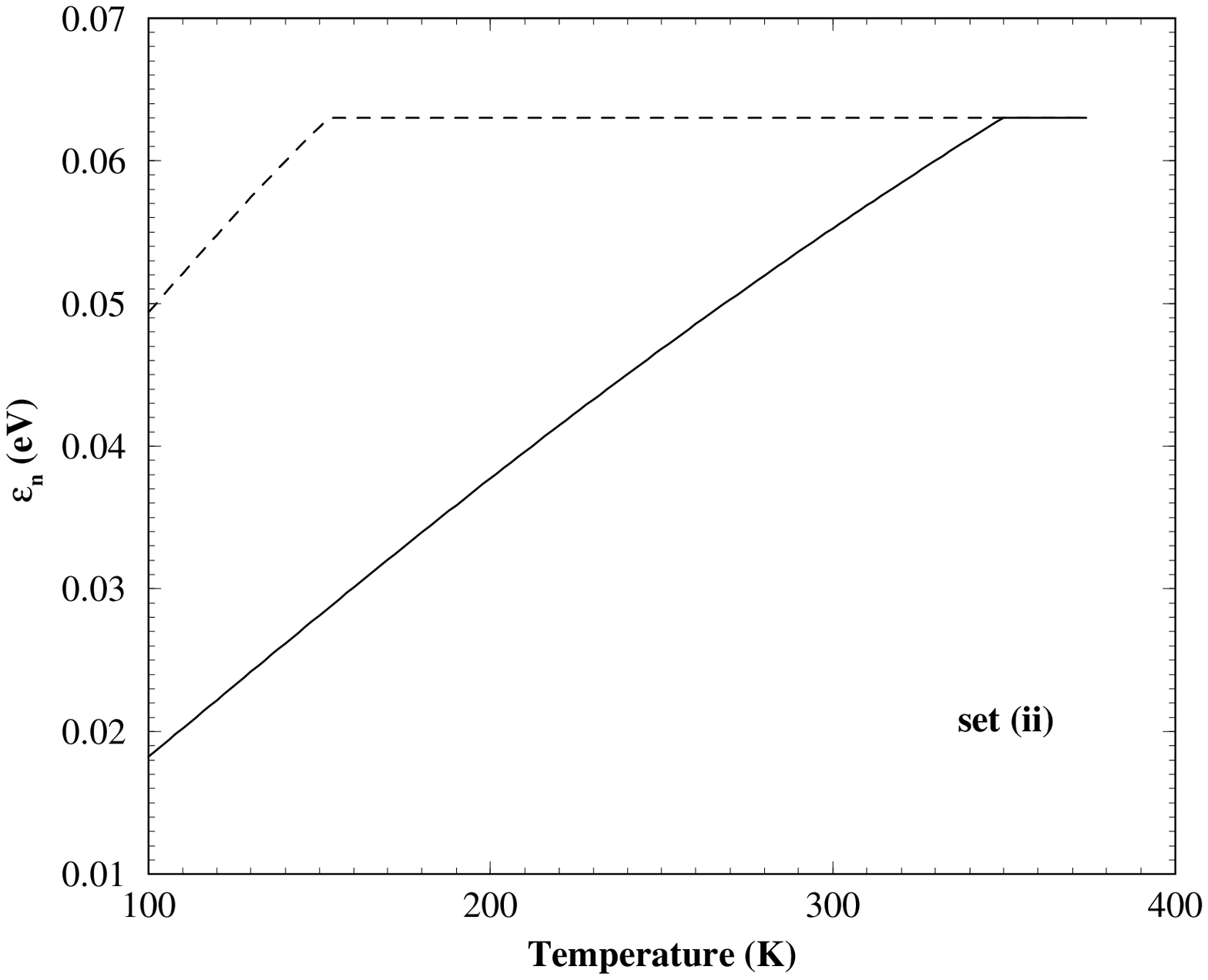} 
\caption{The two lowest eigenvalues $\epsilon_0$ and $\epsilon_1$ of TI (Eq. (9)) 
are plotted as a function of temperature for potential parameters of set (i)  
and set (ii). The temperature at which $\Delta \epsilon = \epsilon_1 - \epsilon_0$ 
becomes zero is the denaturation temperature $T_D$. The value of $T_D$ for the two 
sets of potential parameters is also shown in the figure.}
\label{fig:2}
\end{figure}
In absence of force ($F=0$) and for a homogeneous chain a direct calculation
of the partition function $Z_N^c$ can be performed by the transfer integral
(TI) method \cite{daux95,zhang}. For this it is enough to know the eigenvalues
of the TI;
\begin{eqnarray}
&& a\left(\frac{\beta k}{2\pi a^2}\right)^{1/2} \int dy' 
\exp\left\{-\beta\left[\frac{1}{2}(V(y) +
V(y')) + w(y,y')\right]\right\} \nonumber \\ 
&& \times \phi_i(y')  = e^{-\beta \epsilon_i} \phi_i(y)
\end{eqnarray}
The properties and numerical methods for evaluating the eigenvalues and
eigenfunctions from Eq.(9) have been discussed in details by Dauxois and
Peyrard \cite{daux95} and Zhang et al \cite{zhang}. We have chosen -5.0 ${\rm
\AA}$ and 195.0 ${\rm \AA}$ as lower and upper limits of integration and use
the Gauss-Legendre method to discretize the integral and choose the number
of points $M$ = 900. The eigenvalues and eigenfunctions are found by 
diagonalizing the resulting matrix. The values of
the two lowest eigenvalues $\eps_0$ and $\eps_1$ are plotted in Fig. 2 as a
function of temperature for the two sets of potential parameters. At the
thermal denaturation temperature $T_D$, $\Delta \eps = \eps_1 - \eps_0 \sim 0 $
is minimum. We found that in both cases $\Delta \eps \sim 10^{-5}$ at $T_D$.
The value of $T_D$ found for the potential parameters of set (i) is 350.28 K 
and for the potential parameters of set (ii) 349.66 K. It seems that the 
melting temperature depends on the collective effect of different parts 
of interaction rather than on any one of them.
 
The order of the transition is determined by the exponent $\nu$ which
characterize the gap $\Delta \eps \propto (T_D - T)^{\nu} $ at temperature $T
\leq T_D$; a value $\nu = 1$ implies a cusp in the free energy and a
discontinuous entropy, a feature of first order transition, whereas a value
equal to 2 corresponds to 2nd order transition. Our results plotted in
figures (See Fig. 2) show $\nu = 1$ for both cases. The free energy per base pair is
determined by (see Eq.(6))
\begin{equation}
f = - \frac{1}{2}k_BT \ln \left(\frac{4\pi^2 k_B^2T^2 m}{k}\right) +
\eps_0 \end{equation}
The thermodynamic quantities like the entropy $s$, the specific heat $c_v$ are
evaluated from the standard relations,
\begin{equation}
s = -\frac{\partial f}{\partial T}; \qquad c_v = -T\frac{\partial^2 f}
{\partial T^2}
\end{equation}
A cusp in $f$ at the thermal denaturation temperature $T_D$ is distinctly
seen in the plot of $f$ vs. $T$ in Fig. 3. The cusp lies at the point where
$\Delta \eps = \eps_1 - \eps_0$ becomes zero, i.e. at $T = T_D$. A sharp jump 
in entropy occurs at $T = T_D$. The value of the jump are found to be 
$\Delta s = 3.90\; k_B$ and $2.73\; k_B$, respectively, for potential parameters of 
set (i) and (ii).
This result as well as 
the equilibrium value of base pair stretching $\langle y \rangle $ calculated from 
the relation
\begin{equation}
\langle y \rangle = \int_{-\infty}^{\infty} y |\phi_0(y)|^2 dy
\end{equation}
where $\phi_0(y)$ is the eigenfunction associated with $\epsilon_0$ suggest
that the denaturation transition is first order for both sets of parameters;
though the possibility of underlying continuous transition cannot be ruled out
\cite{barbi}. The apparent first order transition has its origin in the fact
that the thermally generated barrier has a sufficiently larger range than the
Morse potential. A crossover from first order to second order transition is
found for $b/a > 0.5$ \cite{theo}.
\begin{figure}
\includegraphics[height=3.2in,width=3.25in]{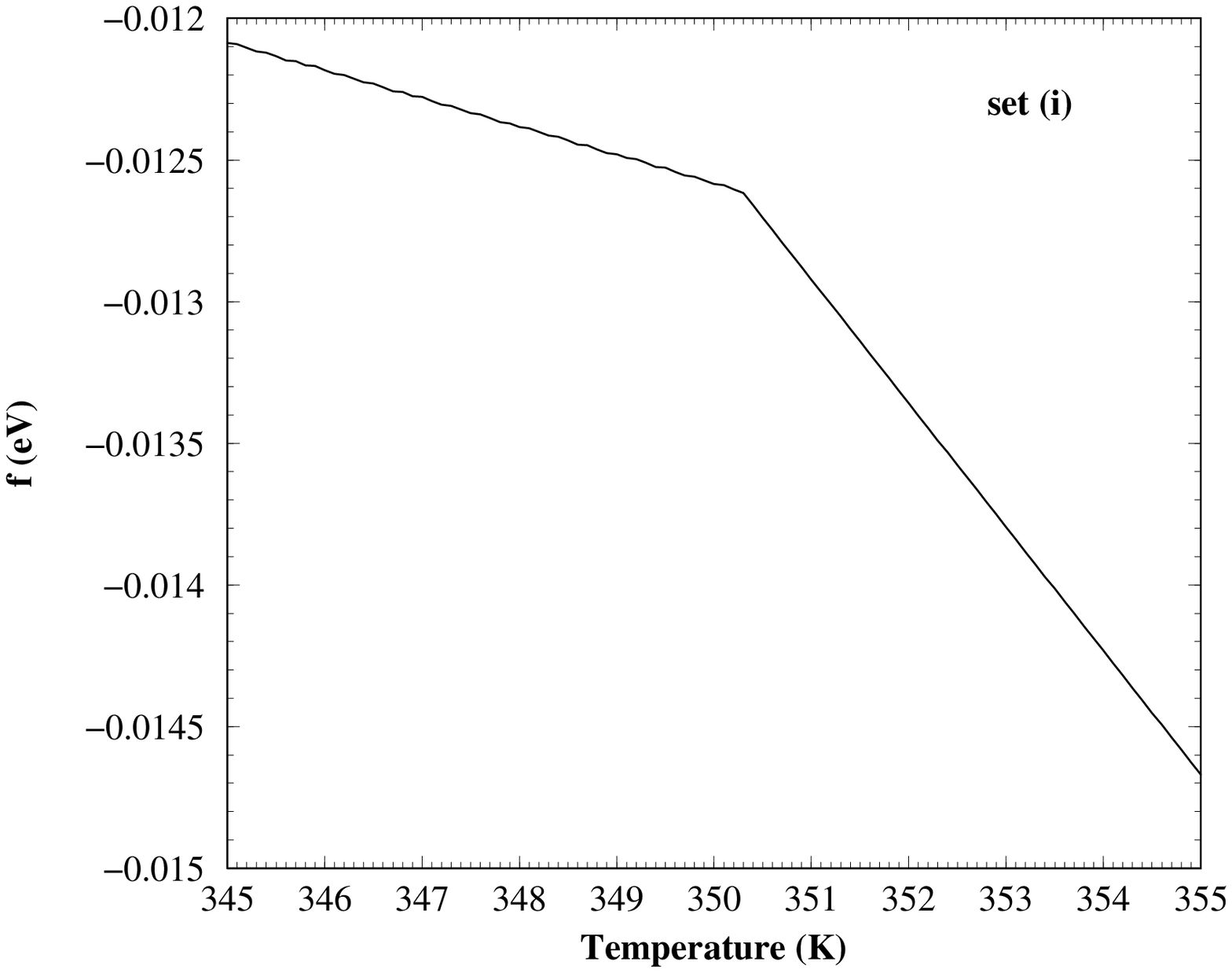} \\
\includegraphics[height=3.2in,width=3.25in]{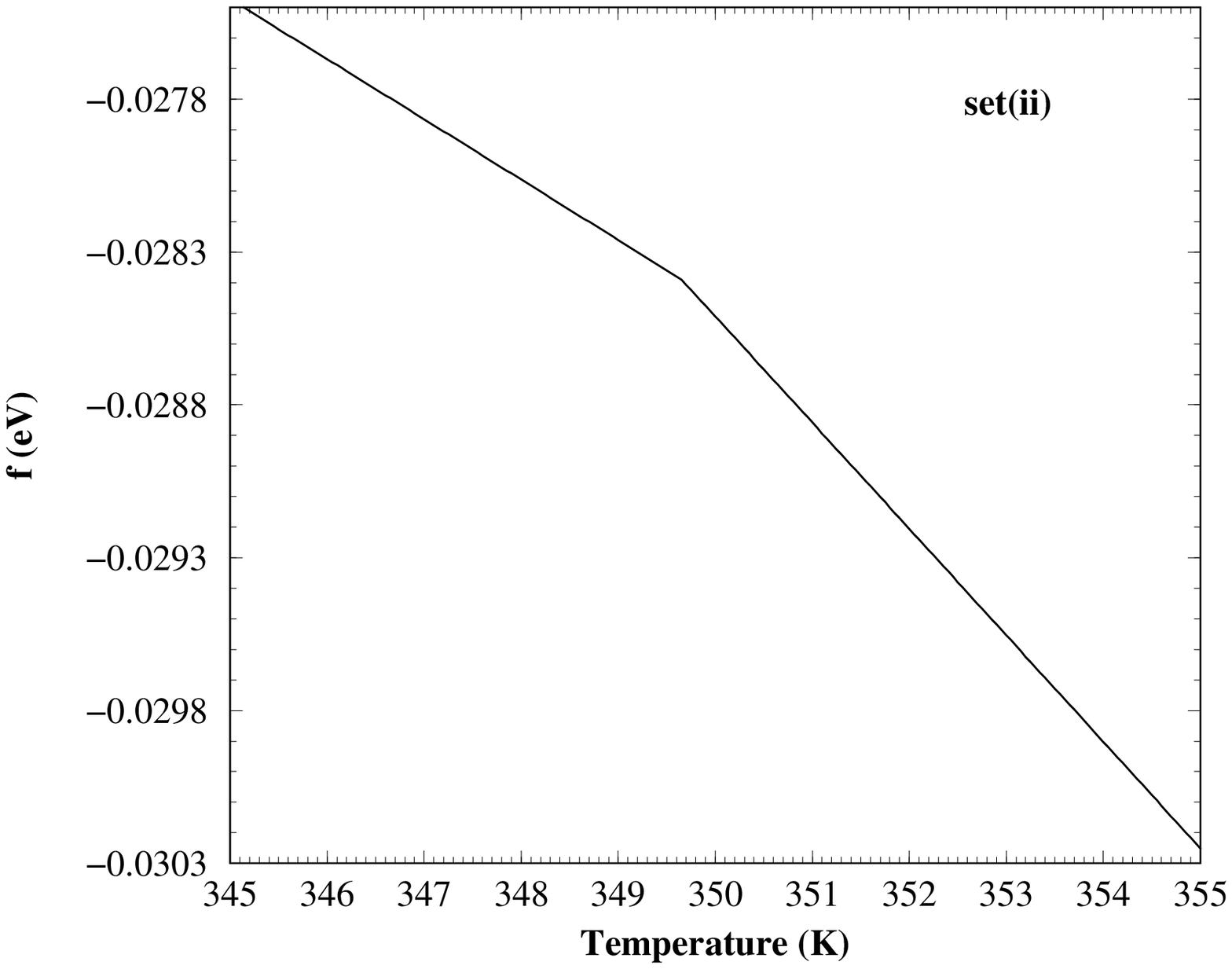} 
\caption{The free energy evaluated using Eq.(10) is plotted as a function of 
temperature for potential parameters of set (i) and set (ii). A cusp in free 
energy at temperature $T = T_D$ is clearly seen. }
\label{fig:3}
\end{figure}

The TI of Eq.(9) can be reduced to a one dimensional Schr{\"o}dinger equation
\cite{cocco2,theo}
\begin{equation}
\left[ -\frac{(k_BT)^2}{2kg(y)}\frac{\partial^2}{\partial y^2} + U(y) \right]
\phi_i(y) = \eps_i \phi_i(y)
\end{equation}
where $g(y) = 1 + \rho e^{-2by}$ and $U(y) = V(y) + (1/2) k_BT \ln
g(y)$.

The solution of Eq.(13) is exactly known when $g(y)$ is replaced by $g(0)$.
For the ground (zipped) state this is a reasonable approximation. Thus we have
\cite{morse}
\begin{equation}
\eps_0 = ak_BT \sqrt {\frac{D}{2kg(0)}} - \frac{(ak_BT)^2}{8kg(0)}
+\frac{1}{2}k_BT \ln g(0)
\end{equation}

The value of $\eps_0$ found from this relation is close to the one calculated
numerically from Eq.(9). The ground state wave function $\phi_0(y)$ for $y \gg 
1/a$ (outside the Morse potential well) can be expressed as
\begin{equation}
\phi_0(y) = \frac{A}{\sqrt {p(y)}} \exp \left[-\int_{y_0}^y dy' \sqrt {p(y')}
\right]
\end{equation}
where $y_0$ is found from the relation 
\begin{displaymath}
U(y_0) = \eps_0 
\end{displaymath}
and 
\begin{displaymath}
p(y) = \frac{1}{k_BT}\sqrt{2kg(y)(U(y) - \eps_0)}
\end{displaymath}

As in the WKB approximation, the coefficient $A$ can be found by connecting
the expression of $\phi_0(y)$ from the two regions in such a way that both the
wave function and its derivative are continuous functions of $y$.

\section{Force Induced Unzipping: Constant Force Ensemble}
\label{sec:4}
When a force $F$ is applied on one end of the ds\d\ as shown in Fig. 1, it
favours the separation of the two strands and one expects a critical force
$F_c(T)$ for unzipping. At a given temperature, $T$, when the applied force
$F$ is less than $F_c(T)$ the \d\ remains in the zipped state and is described
by the lowest eigenvalue $\eps_0$ and eigenfunction $\phi_0(y)$ of Eq.(9).
But for $F > F_c(T)$ the two strands of \d\ will get separated. The free
energy per base pair of unzipped \d\ (i.e. when the particle moves on the
plateau of the Morse potential) can easily be calculated from Eq.(8) as the
Hamiltonian  in this case reduces to
\begin{equation}
H(y_n, y_{n+1}) = D + \frac{1}{2}k(y_n - y_{n+1})^2
\end{equation}
Denoting the free energy per base pair of the unzipped \d\ chain by $g_u$, we
get from Eqs.(8) and (16),
\begin{equation}
g_u = -\frac{k_BT}{N}\ln Z_N^c = D - \frac{F^2}{2k}
\end{equation}
The transition from the zipped to unzipped state takes place when $g_u$
becomes equal to $\eps_0$. Thus the critical force needed to unzip the ds\d\
chain at temperature $T$ is found to be
\begin{equation}
F_c(T) = \sqrt{2k(D-\eps_0)}
\end{equation}

The expression for $F_c(T)$ given by Eq.(18) can also be found using the
following procedure. When the unzipping force $F$ is included in the
expression of the TI of Eq.(9) the one dimensional Schr{\"o}dinger equation
becomes \cite{cocco2},
\begin{eqnarray}
&& \left[ -\frac{(k_BT)^2}{2kg(y)}\frac{\partial^2}{\partial y^2} +
\frac{F(k_BT)}{kg(y)}\frac{\partial}{\partial y} + U(y)
-\frac{F^2}{2kg(y)}\right] \psi_i(y) \nonumber \\
&& = \eps_i \psi_i(y)
\end{eqnarray}
If we substitute the transformation
\begin{equation}
\psi(y) = e^{Fy/k_BT} \phi(y)
\end{equation}
in Eq.(19) it reduces to Eq.(13). Equation(20) suggests that the force
$F$ biases the eigenfunction in the direction of the force. However, as argued
by Lubensky and Nelson \cite{luben} the transformation of Eq.(20) is valid
as long as the eigenfunction $\psi(y)$ satisfies the same boundary
conditions as the eigenfunction $\phi(y)$ and for $\psi_0(y)$ to be well
behaved it is essential that $\phi_0(y)$ should decay at large values of $y$
at least as fast as $\exp(-Fy/k_BT)$.

From Eq.(15) we know that $\phi_0(y)$ decays at large values of $y$ (note
that at large $y$, $g(y) = 1$ and $V(y) = D$ ) as
\begin{equation}
\phi_0(y) \sim \exp\left[-\frac{y}{k_BT}\sqrt{2k(D-\eps_0)}\right]
\end{equation}
The value of force at which the transformation of Eq.(20) breaks down
corresponds to the critical force. Thus from Eqs.(20) and (21) one gets the
expression for $F_c(T)$ given by Eq.(18).

\begin{figure}
\resizebox{0.5\textwidth}{!}{ \includegraphics{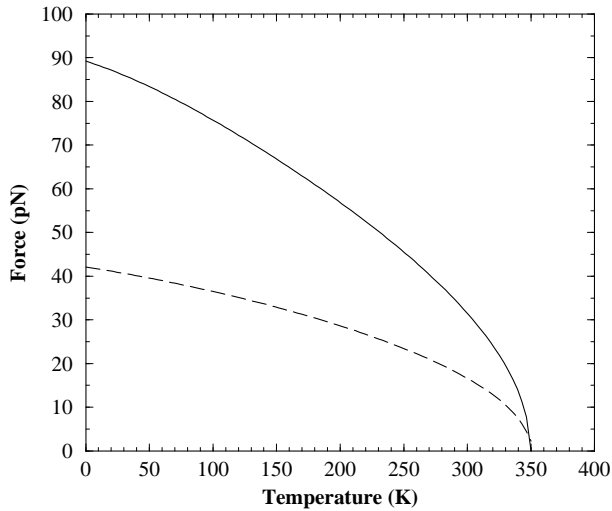} }
\caption{Variation of critical force $F_c(T)$ as a function of temperature for the
two sets of potential parameters. The dashed line corresponds to potential parameters
of set (i) and full line to set (ii).}
\label{fig:4}
\end{figure}
The curve $F_c(T)$ plotted in Fig. 4 gives the boundary that separates the
zipped state of \d\ from the unzipped state in $(T,F)$ plane. We may note that
the value of $F_c(T)$ for the set of potential parameters (i) is less than
that of the force parameters of set (ii) and this difference increases with ($T_D
- T$). This behaviour can easily be understood from the fact that the value of
$F_c(T)$ depends on the product of $k(D-\eps_0)$. Since $D -\eps_0(T)$ is
proportional to $T$ having values equal to $D$ at $T=0$ K and zero at
$T=T_D$, we have
\begin{equation}
F_c(T) = \sqrt{2kD}\left(1-\frac{T}{T_D}\right)^{1/2}
\end{equation}
At a given $T$ the value of $F_c(T)$ depends on $\sqrt{kD}$ as for both sets
of parameters $T_D$ are nearly equal. This explains the difference in the
value of $F_c(T)$ found for the two sets of parameters. Equation (22) is
important as it gives the dependence of $F_c(T)$ on the stiffness of single
strand, the potential well depth of interaction in a base pair and on the
temperature.

It may, however, be noted that the experimental verification of Eq.(18)
or (22) requires that the forces that stabilize the ds\d\ structure and
have been included in the Hamiltonian (see Sec. II) be temperature
independent and the separated strands of \d\ do not form hair-pin or
globule like structures \cite{danil1}.

\section{Constant Extension Ensemble}
\label{sec:5}
\subsection{Extension of one of the ends base pair}
\label{sec:6}
One can devise experiments in which the separation of one end of the two
strands of ds\d\ is kept fixed and the average force needed to keep this
separation can be measured. In fact the experimental set up of
Essevaz-Roulet \cite{roulet} to unzip \d\ by displacing the bead at
constant velocity belongs to the category of the constant extension ensemble.

To calculate the force needed to keep one of the ends base pair of
the chain to a given separation let us consider a chain of $N$ base
pairs of which the base pair 1 is stretched to a distance $y$ and
the $N^{th}$ base pair is fixed to zero separation ({\it i.e.} $y_N = 0$).
The configurational partition function (see Eq. 6) of this
chain is
\begin{eqnarray}
Z_N^c(y) &=& \left( \frac{\beta k}{2\pi}\right)^{N/2} \int
\prod_{n=1}^{N-1} dy_n \delta(y_1 - y) \delta(y_N - 0) \nonumber \\
& \times &\exp\left\{-\beta\sum_{n=1}^{N-1}\left[\frac{1}{2}(V(y_n) + V(y_{n+1}) )
+ w(y_n,y_{n+1}) \right]\right\} \nonumber \\
&&
\end{eqnarray}
In terms of the eigenvalues and eigenfunction of transfer integral
operator defined in Eq. (9) we can write Eq. (23) as
\begin{eqnarray}
Z_N^c(y) &=& e^{-\beta V(y)/2} \phi_0(y) e^{-\beta(N-1)\epsilon_0}
\phi_0(y_N = 0) e^{-\beta V(y_N = 0)/2} \nonumber \\
&=& C e^{-\beta V(y)/2}\phi_0(y)Z_N^c
\end{eqnarray}
where the constant $C =  e^{\beta(\epsilon_0 - D/2)}\phi_0(y_N = 0)$
and $Z_N^c = e^{-\beta N \epsilon_0}$.

The work done in stretching the first base pair to distance $y$ is
therefore (as the constant $C$ is $y$ independent, we drop it henceforth)
\begin{subequations}
\begin{eqnarray}
W(y) & = &  \frac{1}{2}V(y) - k_B T \ln \phi_0(y) \\
& = & \frac{1}{2}V(y) - k_BT [\ln Z_N^c(y) - \ln Z_N^c] 
\end{eqnarray}
\end{subequations}
The derivative of $W(y)$ with respect to $y$ gives the average force $F(y)$ that is
needed to keep the extension equal to $y$. Thus,
\begin{equation}
F(y) = \frac{\partial W(y)}{\partial y}
\end{equation}

In Eqs. (24) \& (25) $V(y)/2$ appears as an end term. This is because in constructing
the transfer integral that connects the base pair $1$ with the base pair
$2$ only $V(y)/2$ has been taken into account and remaining $V(y)/2$
acts as an end term. 
When the periodic boundary condition is imposed as in the study of thermal 
denaturation, the term $V(y)/2$ of base pair 1 and the term $V(y_N)/2$
of base pair $N$ get absorbed in the transfer integral that connects the base
pair 1 with the base pair $N$ and therefore no end term appears.

For a homogeneous chain one can use the eigenfunction $\phi_0(y)$ found by solving
Eq. (9) to calculate $W(y)$ from Eq. (25a). Alternatively, one can use the method
of matrix multiplications to calculate the partition functions $Z_N^c(y)$ and
$Z_N^c$ and use Eq. (25b) to find the value of $W(y)$. The method of matrix
multiplication is useful as it can be applied to cases ({\it e.g.} heterogeneous chains) 
where the method of transfer integral is not applicable.

In the method of matrix multiplication one first constructs a matrix for each
base pair by using the potential parameters corresponding to the base pair under 
consideration. The discretization of the coordinate variables and introduction
of a proper cutoff on the maximum value of $y's$ determine the size of the 
matrices and the number of base pairs in the chain the number of matrices
to be multiplied \cite{nav,campa}. As in evaluation of eigenvalues and eigenfunctions 
from Eq.(9)
we chose -5.0 and 195.0 ${\rm \AA}$ as the lower and upper limits of integration
and the Gauss-Legendre method to discretize the integral. We found that the number
of grid point $M=450$ leading to 450 $\times$ 450 matrix for each base pair
gives good result. We have considered  chains of number of base pairs varying
from 100 to 300 and compared their results in Table 1 and 2 at $T = 200$ K
and $300$ K, respectively for the potential parameters of set (ii). 
In Fig. 5 we compare the result found for $W(y)$ and $F(y)$ for a chain 
of 100 base pairs with that of the result found from Eq.(25a) using the 
eigenfunctions $\phi_0(y)$. In view of excellent agreement found between the
values calculated from the two methods we conclude that even a chain of 100
base pairs is good enough to calculate $W(y)$ and $F(y)$ and some other
related properties (as discussed below) where the method of transfer integral
is not valid. In such calculations one has, however, to be careful to keep
the value of $y$ sufficiently small compared to length of the molecule.
\begin{figure}
\includegraphics[height=4.5in,width=3.5in]{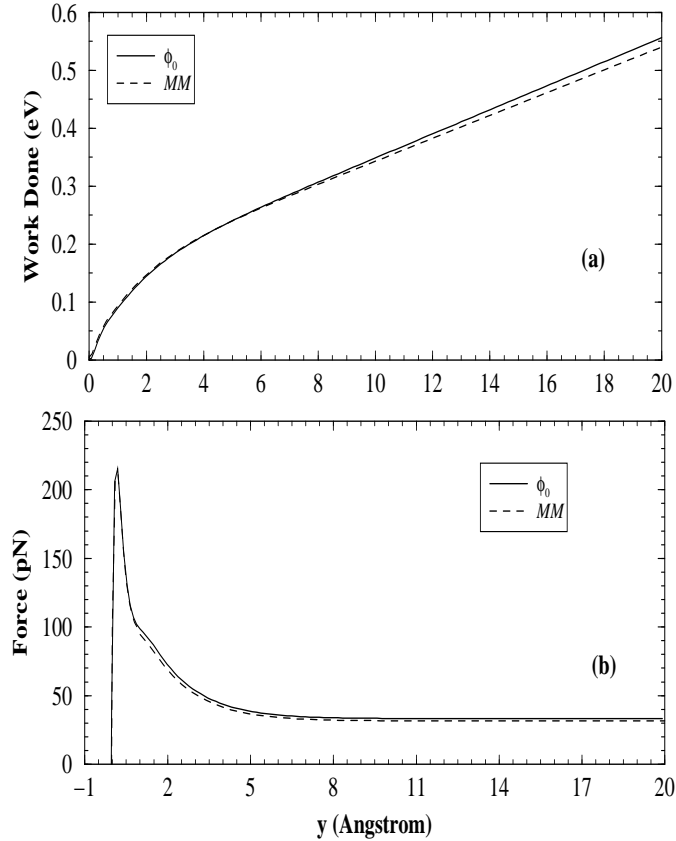} 
\caption{Comparision of (a) work done and (b) force needed to keep end base pair
at a distance $y$ calculated using groundstate eigenfunction and 
matrix multiplication methods (Eqs. 25a \& 25b). This curve is at $T$ = 300 K.}
\label{fig:4}
\end{figure}

The force $F(y)$ calculated using Eq. (26) is shown in Fig. 6 for both sets
of potential parameters at $T = 300$ K. We note that the existence of a very
large force barrier at short distance ($y \sim 0.2 \; {\rm \AA}$) and large
difference in the values of the force at small extension from the two
sets of potential parameters. For example, the value of the force at the
peak for the set of parameters (i) is about thrice that
found for the set of parameters (ii). The width of the peak for the set of 
parameters (i) is also about thrice that of the set of parameters (ii).
\begin{table}[h]
\caption{Values of work $W(y)$ done in stretching one of the ends base pair
of ds\d\ molecule of $N$ base pairs to a distance $y$ at $T$ = 200 K for potential
parameters of set (ii). }
\begin{tabular}{|c|c|c|c|c|} \hline 
&\multicolumn{4}{c} {Work done (eV)} \vline \\ \hline 
$N$ & $y = 0.5 \; {\rm \AA}$ & $y = 1.0 \; {\rm \AA} $ & 
$y = 5.0 \; {\rm \AA} $ & $y = 10.0 {\rm \AA} $ \\ \hline
100 & 0.0154 & 0.0397 & 0.2470 & 0.4296  \\
300 & 0.0154 & 0.0397 & 0.2470 & 0.4296  \\ \hline
\end{tabular} 
\end{table}

\begin{table}[h]
\caption{Values of work $W(y)$ done in stretching one of the ends base pair
of ds\d\ molecule of $N$ base pairs to a distance $y$ at $T$ = 300 K for potential
parameters of set (ii). }
\begin{tabular}{|c|c|c|c|c|} \hline 
& \multicolumn{4}{c} {Work done (eV)} \vline \\ \hline
$N$ & $y = 0.5 \; {\rm \AA}$ & $y = 1.0 \; {\rm \AA} $ & 
$y = 5.0 \; {\rm \AA} $ & $y = 10.0 {\rm \AA} $ \\ \hline
100 & 0.0136 & 0.0327 & 0.1769 & 0.2801 \\
300 & 0.0136 & 0.0327 & 0.1769 & 0.2801 \\ \hline
\end{tabular}
\end{table}
\begin{figure}
\resizebox{0.5\textwidth}{!}{ \includegraphics{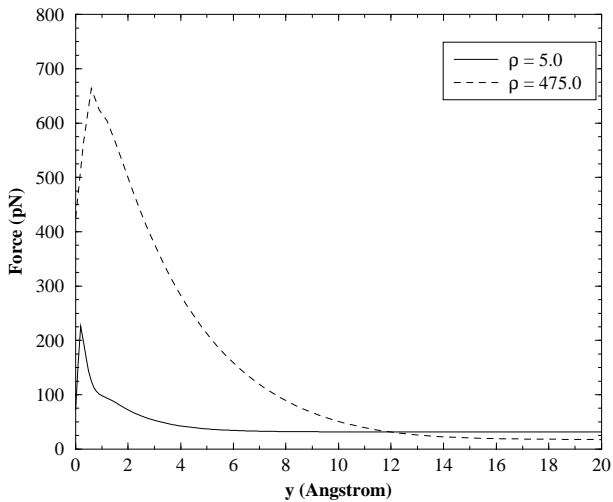} }
\caption{The average force $F(y)$ in pN required to stretch one of the ends base
pair to a distance $y$ at $T$ = 300 K. The minimal separation $y$ = 0 corresponds
to the ds\d\ equilibrium structure. For sufficiently large value of $y$ the
$F(y)$ approaches to the value found from the constant force ensemble and
shown in Fig. 4. The dashed line corresponds to potential parameters of set (i) and
full line to set (ii).} 
\label{fig:5}
\end{figure}

As suggested in ref.\cite{cocco2} the physical origin of the large force
barrier is in the potential well due to hydrogen bonding plus the additional
barrier associated with the reduction in \d\ strand rigidity as one passes
from ds\d\ to ss\d . The large difference in the value of the force on the
peak for the two sets of potential parameters is primarily due to large
difference in the barrier associated with the reduction in the \d\ strand
rigidity. In other words, the value of the force at the peak depends rather
sensitively on the value of $\rho$ and the value of the potential depth $D$.
Beyond $y \sim 1.0 \;{\rm \AA}$ the contribution arising due to the 
potential $V(y)$ becomes almost negligible compared to the contribution
arising due to the term involving $\phi_0(y)$. 
\begin{figure}
\resizebox{0.45\textwidth}{!}{ \includegraphics{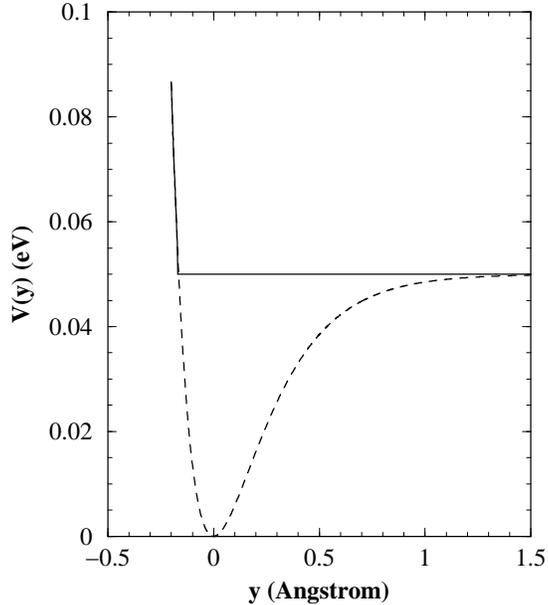} }
\caption{The on-site potential $V(y)$ as a function of displacements. The dashed 
line represents the Morse potential (see Eq. (4)) and the full line the 
potential chosen to represent the interaction at the defect sites.}
\label{fig:6}
\end{figure}

The single and double stranded portions of the molecule are separated by a
boundary region. In this boundary region which may be of three or four base
pairs length \cite{cocco2} the bases are unpaired but not free to fluctuate.
The peak corresponds to the energy needed to create this boundary. To test
validity of this argument we calculated $F(y)$ curve by replacing first
few base pairs by defects. A defect on a \d\ chain means mismatched base pair
\cite{nav}. For example, if one strand of \d\ has adenine on a site, the other
strand has guanine or cytocine or adenine in place of thymine on the same
site. In such a situation the pair will remain in open state at all
temperatures as the two nucleotide cannot join each other through hydrogen
bonds. We therefore replace the on-site Morse potential by a potential shown
in Fig. 7 by full line. This potential has repulsive part as well as the flat
part of the Morse potential but not the well that arises due to hydrogen
bonding interactions. Due to defect on a site the stacking interactions with
adjacent bases will also be affected. Since the formation of hydrogen bonds
changes the electronic distribution on base pairs causing stronger stacking
interactions with adjacent bases. Therefore, when base pairs without hydrogen
bonds are involved the stacking interaction will be weaker compared to the
case when both base pairs are intact. This fact has been taken into
account in our calculation by reducing the anharmonic coefficient $\rho$ to
its half value whenever one of the two base pairs involved in the stacking
interactions is defective and zero when both are defective.

For given number of defect base pairs we calculated $W(y)$ and $F(y)$ using
Eq.(24) and (25). The results are shown in Fig. 8 for $T = 200 $ K and
in Fig. 9 for $T = 300 $ K.
\begin{figure}
%\resizebox{0.5\textwidth}{!}{ \includegraphics{fig8.eps} }
\includegraphics[height=4.5in,width=3.25in]{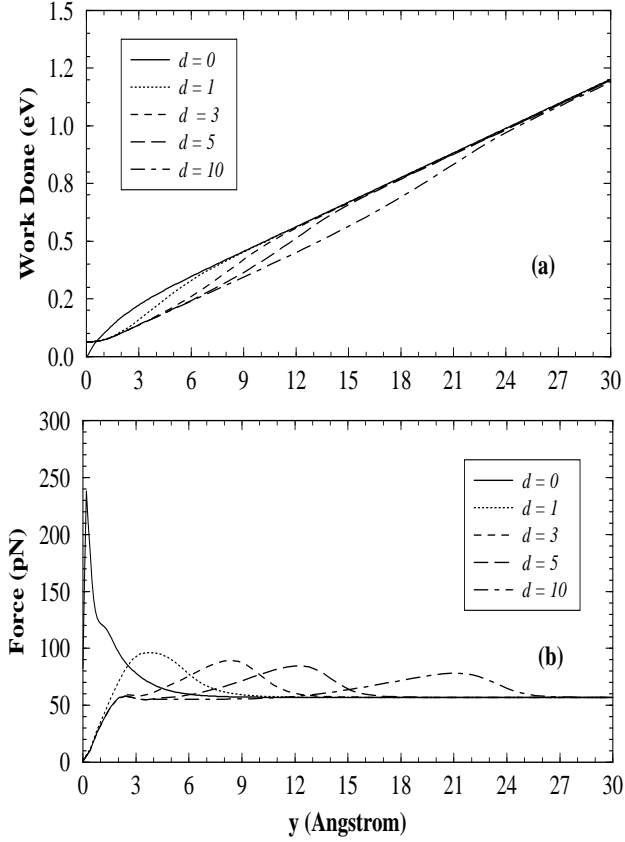} 
\caption{(a) The work $W(y)$ in eV done at $T$ = 200 K in stretching one of
the ends base pair to a distance $y$ and the change in the value of $W(y)$
when some of the base pairs including the one being stretched are replaced by 
defect base pairs are shown. In (b) the force $F(y)$ in pN at $T$ = 200 K required
to stretch one of the ends base pair to a distance $y$ in presence of defect
base pairs is shown. The peak position is found to shift to larger values of 
$y$ as the number of defect base pairs is increased. The results plotted
in this figure are obtained using the potential parameters of set (ii).}
\label{fig:7}
\end{figure}
\begin{figure}
%\resizebox{0.5\textwidth}{!}{ \includegraphics{fig9.eps} }
\includegraphics[height=4.5in,width=3.25in]{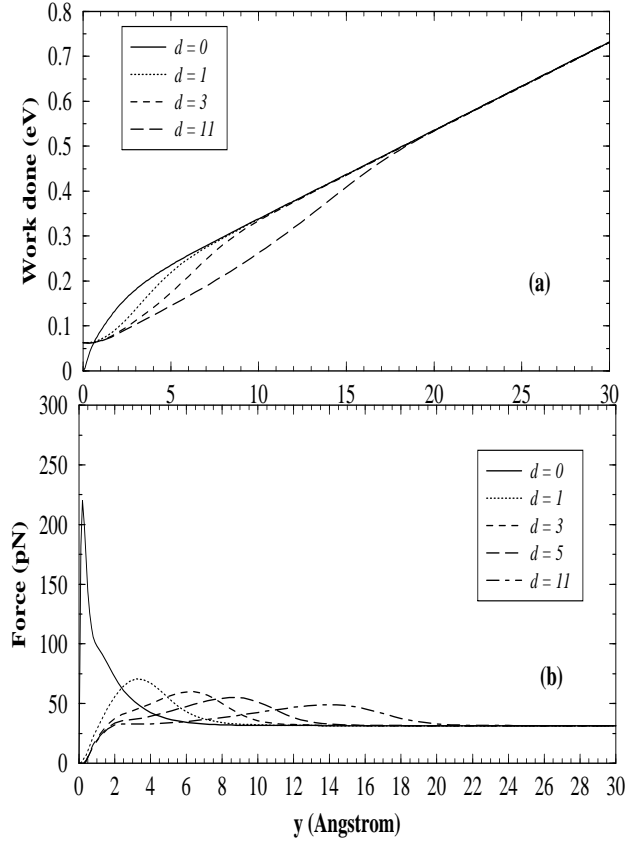} 
\caption{The curves in (a) and (b) are same as in Fig. 8 but for $T$ = 300 K.}
\label{fig:8}
\end{figure}

As expected, the work needed to create a separation $y$ of an end base pair
decreases as the number of defect base pairs increases. When $y$ becomes
larger than certain value which, in turn, depends on the number of defect base
pairs (see Figs. 8 \& 9) $W(y)$ becomes equal to that of a defectless ds\d .
It also seems that for small $y$ the value of $W(y)$ attains a lower limit
that is independent of the number of defect base pairs. This behaviour can
also be seen in $F(y)$ curve; as for small $y$ the value of $F$ is (except
with no defect) almost same for all cases and for large $y$ the value of
force approaches to its asymptotic value. It is interesting to note that the
peak in $F(y)$ curve shifts to larger values of $y$ as the number of
defect base pairs increases and the height of the peak decreases. These
features can be understood from the fact that the barriers that give rise the
peak in $F(y)$ curve shift to the base pairs after
the segment of the \d\ that contains defect base pairs. Note that this segment
of \d\ with defect base pairs is in single stranded form and can have
comparatively large thermal fluctuations. The energy associated with this
fluctuation (entropic) contributes to reducing the barrier that is
responsible for the peak in $F(y)$ curve. As the length of this defect
segment increases the entropic contribution increases and therefore more
reduction in the height of the peak in $F(y)$ curve.

Since the effect of temperature is to reduce the effective height of
barriers responsible for the peak and to increase the
entropic contributions, the peak height in $F(y)$ curve are smaller at $T =
300$ K compared to that at $T = 200 $ K.

In Table 3 we list the value of force at the peak and the peak position for a
number of defect base pairs calculated using the potential parameters of set
(ii). We note that the position of the peaks at $T = 200$ K are at
larger values of $y$ compared to the corresponding peaks at $ T = 300 $ K.
For example, for the first five base pairs being defective, the peaks are
found for $y = 12.4 {\rm \AA}$ and $8.8 {\rm \AA}$, respectively at $T = 200
$ K and $300$ K. This means, one has to have larger extension to encounter the
force peak in $F(y)$ curves at lower temperatures compared to that found at
higher temperatures. This can be understood from the fact that (as shown
below) the extension at one end in a ds\d\ creates a fork of Y shape and
the length of this fork increases as extension $y$ increases. As we will see
that to have same length of the fork one has to have larger extension at low
temperature compared to that at high temperatures. Therefore to reach to those
base pairs which are responsible for the barriers leading to peak in $F(y)$
one has to have relatively larger extension as temperature is lowered.

\begin{table}[h]
\caption{Peak position and the value of force at the peak 
for different number of defect base pairs located at one end of ds\d .}
\begin{tabular}{|c|c|c|c|c|} \hline
& \multicolumn{2}{c} {T = 200 K} \vline &  \multicolumn{2}{c} {T = 300 K} \vline
\\ \hline
$N_d$ & P. Position  & P. Height  & P. Position & P. Height \\ \hline
 & (in {\rm \AA}) &( in pN) & (in {\rm \AA}) &( in pN)\\ \hline
0   & 0.19  & 235.77 & 0.19  & 226.48  \\
1   & 3.6  & 96.33  & 3.2  &  70.57  \\
3   & 8.4  & 89.53  & 6.4  &  59.90  \\
5   & 12.4 & 84.60  & 8.8  &  55.04  \\
11  & 22.8 & 77.22  & 14.0 &  49.11 \\ \hline
\end{tabular} 
\end{table}
To see the formation of a fork of Y shape on stretching the two
strands of a ds\d\ from one end we calculate $\langle y_n \rangle$ for $n > 1$
for the value of $y_1 = y$ from the relation
\begin{equation}
\langle y_n \rangle = \frac{\int\left(\prod_{i=1}^N dy_i\right) y_n
\exp[-\beta \sum_{i=1}^N H(y_i, y_{i+1})]\delta(y_1-y)}{\int \left(\prod_{i=1}^N dy_i
\right) \exp[-\beta \sum_{i=1}^N H(y_i, y_{i+1})]}
\end{equation}
We use matrix multiplication method to find $\langle y_n \rangle$ for a ds\d\
of 200 base pairs. We have checked the accuracy of our results given below by
using ds\d\ of longer sizes and found that as long as $y$ is kept
sufficiently small compared to the size of the chain, results remain
independent of the size of the chain. In Fig. 10 we plot the values of
$\langle y_n \rangle$ showing the average position of the two strands at
different values of extension. The formation of Y fork at the end being
stretched is clearly seen. The length of the fork increases on increasing the
value of $y$. We also note that the effect of temperature on the length of
the fork. For example, at $y = 5 {\rm \AA} $ the junction of the two
branches of the fork is located at base pairs $n = 4 \; {\rm and} \; 6$
respectively at $T = 200 \; {\rm and} \; 300 $ K.
\begin{figure}
%\resizebox{0.5\textwidth}{!}{ \includegraphics{fig10.eps} }
\includegraphics[height=4.0in,width=3.25in]{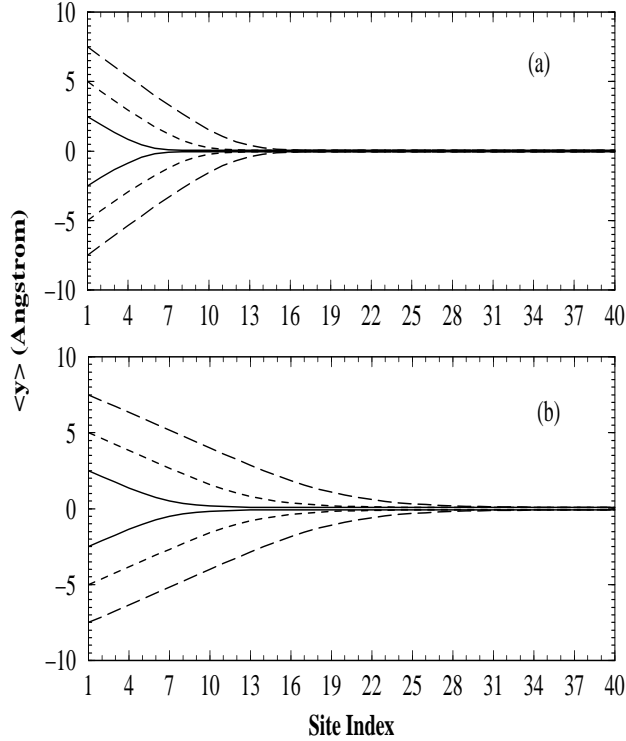} 
\caption{The shape and size of fork of Y shape formed at $T$ = 200 K (a)
and $T$ = 300 K (b) when one of the ends base pair is stretched. $\langle y \rangle $
measures the separation in angstrom of the two strands at different sites (base pairs)
numbered from 1 to $N$.}
\label{fig:9}
\end{figure}
We consider $n^{th}$ base pair open if $\langle y_n \rangle$ is equal or
greater than $1{\rm \AA}$ and bound or intact if value of $\langle y_n
\rangle$ is less than $1{\rm \AA}$. In Fig. 11 we plot the number of open base
pairs as a function of extension $y$. We note that except for very small
values of $y$ the number of open base pairs increases linearly with the
extension and the slope of the line corresponding to 300 K is about twice
as compared to that for 200 K.
\begin{figure}
\resizebox{0.5\textwidth}{!}{ \includegraphics{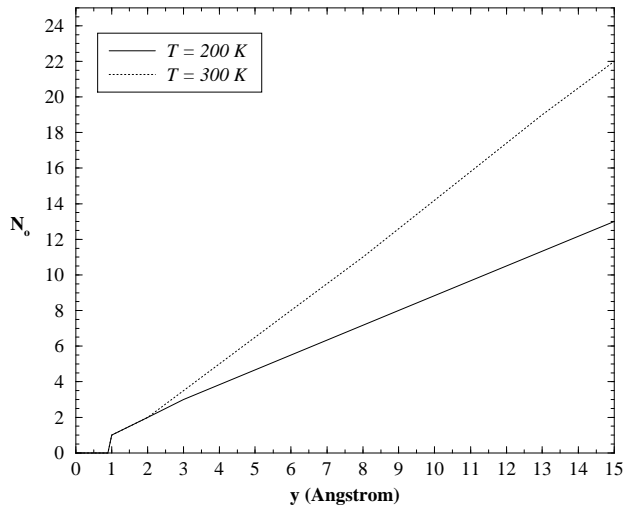} }
\caption{The number of open base pairs, $N_o$, as a function of the extension of the
end base pair at temperature $T$ = 200 K (full line) and 300 K (dashed line).
A base pair is considered to be open when the value of $\langle y \rangle $ is 
equal or greater than 1 ${\rm \AA}$ and closed if it is less than 1 ${\rm \AA}$. 
At initiation of unzipping the number of base pairs that
get open depends on the anharmonic term in the stacking interaction (see Eq. 5)
but after some extension it depends linearly on $y$. The slope of the curve
at $T$ = 300 K is about twice the corresponding value at $T$ = 200 K.}
\label{fig:10}
\end{figure}

\subsection{Extension of a base pair in the middle}
\label{sec:7}
In replication the opening of ds\d\ is initiated at one of the ends whereas
in case of the transcription it can be anywhere. It is therefore of
interest to investigate the formation of bubble of ss\d\ in a ds\d\ away
from the ends.

We consider the situation in which a bubble of ss\d\ is formed in the
middle of a ds\d\ by stretching a base pair and calculate the average
force needed to create it. For this we use the method of matrix
multiplication described above and calculate the work $W(y_m)$ done
in stretching the middle base pair by a distance $y_m$ from the relation
(see Eq.(24))
\begin{equation}
W(y_m) = -k_BT (\ln Z_N^c(y_m) - \ln Z_N^c)
\end{equation}
where $Z_N^c(y_m)$ is the partition function of ds\d\ of $N$ base pairs
with middle base pair kept at $y_m$  separation. For our calculation we
have taken $N = 200$ as we pointed out earlier that as long as $y_m$ is
small the results are independent of length of the molecule and constructed the
matrix using the procedure already described.
\begin{figure}
%\resizebox{0.5\textwidth}{!}{ \includegraphics{fig12.eps} }
\includegraphics[height=4.5in,width=3.25in]{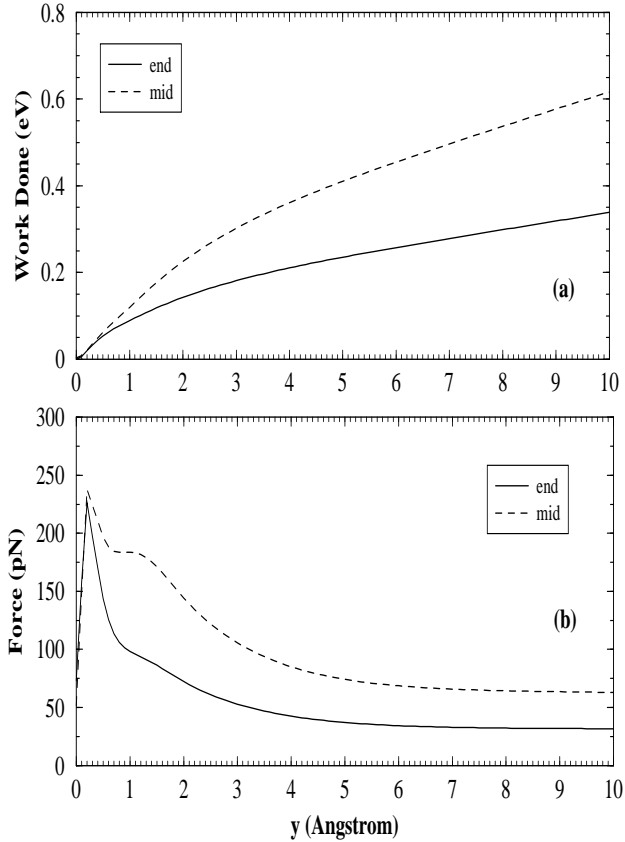} 
\caption{Comparison of (a) the work $W(y)$ that has to be done and (b) the average
force $F(y)$ needed when a base pair of one of the ends or in the middle of the 
molecule is stretched to a distance $y$. The results plotted here correspond to 
potential parameter of set (ii) at $T$ = 300 K}
\label{fig:11}
\end{figure}

The values of $W(y)$ and $F(y)$ found from this calculation are plotted
in Fig. 12 and as shown in the figures the values for any given extension
are exactly twice the values when one of the ends base pair
was stretched only when $y \geq 1.0 \; {\rm \AA}$ but not for $y < 1.0 {\rm \AA}$. 
This is because for the extension $y < 1.0 \; {\rm \AA}$ a contribution
due to $V(y)/2$ arises in the case of extension of one of the ends base pair
but not in the case of extension of a base pair away from the ends. The reason
why one gets for $y\geq 1.0 \; {\rm \AA}$ the force needed to stretch a base
pair in the middle twice that of the base pair at one ends is due to
the fact that a bubble of open base pairs formed in the middle has to
propagate on both sides in contrast to the earlier situation in which it
has to move in one direction only. 

Alternatively, we can use the argument used
in writing Eq.(25a). As the probability of finding the middle (or for that
matter any base pair away from the ends) at a separation of $y$ is
$|\phi_0(y)|^2$, the work done in achieving the extension $y$ is, therefore,
\begin{equation}
W(y) = -2k_BT\ln\phi_0(y)
\end{equation}
Note that this term is twice the second term in Eq. (25a) only. 
Since the contribution arising due to first term in Eq. (25a) 
is only for small values of $y$ (i.e. $y < 1.0 \; {\rm \AA}$), the work done
in pulling a base pair that is far away from the ends, is twice the work done
in pulling one of the ends base pair for $y \geq 1.0 \; {\rm \AA}$.
The asymptotic value of the force which is equal to
the one found from the constant force ensemble is, however, exactly two
times to the value corresponding to the extension done at the one end.
\begin{figure}
%\resizebox{0.5\textwidth}{!}{ \includegraphics{fig13.eps} }
\includegraphics[height=4.5in,width=3.25in]{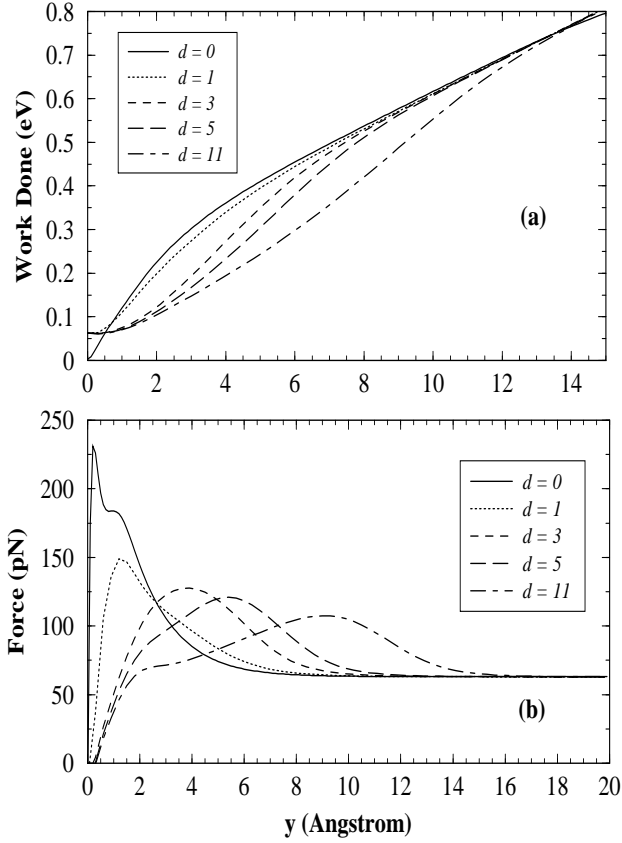} 
\caption{The curves in (a) and (b) are same as in Fig. 9 with a difference that 
instead of a base pair at the end, a middle base pair of a ds\d\ molecule is 
stretched and defect base pairs are introduced symmetrically about the middle 
base pair. The results plotted here are for $T$ = 300 K and correspond to the
potential parameter of set (ii). }
\label{fig:12}
\end{figure}

The effect of defects on $W(y)$ and $F(y)$ has been calculated by making
the base pair that is being stretched and others symmetrically on both
sides of it as defect base pairs. The result found for different number
of defect base pairs are shown in Fig. 13 for $T = 300$ K. Comparing the
results of this figure with those of Fig. 9 we find that the qualitative
nature of $W(y)$ and $F(y)$ curves in these two cases is similar except in the
case when one base pair (i.e. the base pair which is being stretched) is a
defect base pair. It can be seen that the peak position in $F(y)$ in Fig. 9 
(and Table 3) is shifted from 0.19 ${\rm \AA}$ to 3.2 ${\rm \AA}$ whereas in
the case shown in Fig. 13 there is very little shift.
This means that in the latter case the position of the barrier that gives rise
the peak in $F(y)$ does not shift. The decrease in the value of force at the
peak is primarily due to loss of energy of hydrogen bonds in the base pair.
The other important point to be noted is that in the presence of defects the
peaks in $F(y)$ is more than twice the corresponding peak in Fig. 9.
This can be understood from the fact noted above that decrease in the height
of peak is due to contribution arising from thermal fluctuations in the
segment of \d\ containing defects. This contribution is larger when the
segment is located at the end of the chains than in the middle. In Table 4, we
list the position and height of the peak in $F(y)$ curve of these two cases
for $T = 300$ K. We note that the peaks occur at smaller values of extension
compared to the case of end extension.
\begin{table}[h]
\caption{Comparison of the peak position and the values of force at the
peak for different number of defect base pairs introduced at one end
and in the middle of ds\d\ at $T$ = 300 K.}
\begin{tabular}{|c|c|c|c|c|} \hline
& \multicolumn{2}{c} {Defect in middle} \vline &
\multicolumn{2}{c} {Defect at end} \vline
\\ \hline
$N_d$ & P. Position & P. Height & P. Position & P. Height \\ \hline
& (in {\rm \AA}) & (in pN) & (in {\rm \AA}) &  ( in pN)\\ \hline
0   & 0.19  & 239.43  & 0.19 &  226.48  \\
1   & 1.25 & 148.77  & 3.2  &  70.57  \\
3   & 5.1  & 127.67  & 6.4  &  59.90  \\
5   & 6.3  & 120.95  & 8.8  &  55.04  \\
11  & 8.7  & 107.52  & 14.0 &  49.11  \\
15  & 12.0 & 102.33  & 19.2 &  46.13  \\ \hline
\end{tabular}
\end{table}

The value of $\langle y_n \rangle$ calculated from Eq.(26) with the
modification that now the middle base pair is kept at fixed value are plotted
in Fig. 14. Formation of an ``eye shape'' bubble in the middle is clearly seen.
The length of the bubble increases symmetrically as the extension of the
middle base pair is increased.
\begin{figure}
%\resizebox{0.5\textwidth}{!}{ \includegraphics{fig14.eps} }
\includegraphics[height=2.5in,width=3.25in]{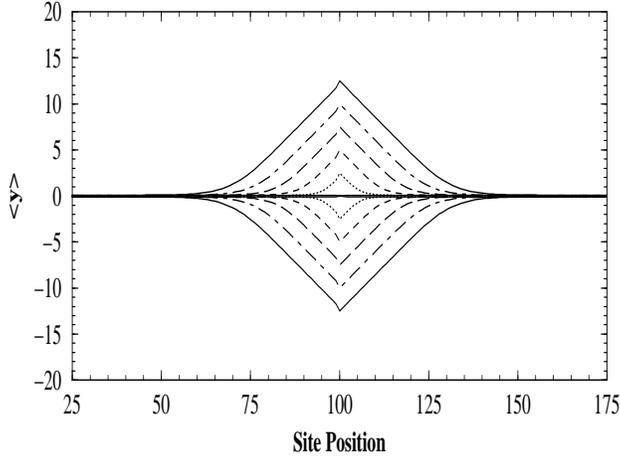} 
\caption{ Formation and elongation of a bubble of ss\d\ with a shape of an 
eye at $T$ = 300 K when a base pair far away from  the ends is stretched. The
results plotted corresponds to potential parameters of set (ii) at $T$ = 300 K.}
\label{fig:13}
\end{figure}

\section{Summary and Conclusion}
\label{sec:8}
The basic features of unzipping of a ds\d\ molecule under the influence of
external force have been investigated using the PB model. The model, though
ignores the helicoidal structure of ds\d\ molecule, has enough details to
analyze the mechanical response at the few ${\rm \AA}$ scale and is simple
enough for numerical and analytical analysis.

The critical force $F_c(T)$ for unzipping calculated in the constant force
ensemble is found to depend on the potential parameters $k$ and $D$ measuring,
respectively, the stiffness of a single strand of \d\ and the depth of the
on-site potential. The temperature dependence of $F_c(T)$ is found to be
$(T_D - T)^{1/2}$ where $T_D$ is the thermal denaturation temperature in
absence of the external force. It is, however, important to note that our
approach assumes that the interactions that stabilize the ds\d\ structure and
are included in the PB model are temperature independent and separated strands
of \d\ do not fold to form hair-pin or globule-like structures. Both these
assumptions may not strictly be met in systems commonly used in experiments
\cite{danil1}. Therefore, experimental verification of Eq.(22) needs special
care.

In the constant extension ensemble the average force $F(y)$ needed to stretch
a base pair $y$ distance apart is found to have a large barrier at the
separation of the order of $0.2 {\rm \AA}$. A similar result has been reported
by Cocco et al \cite{cocco2}. The value of $F(y)$ for any $y$ depends on
whether the base pair being stretched is one of the ends base pair or located
far away from the ends of the molecule. The value of $F(y)$ of the latter case 
for a given extension $y$ is found exactly twice of the value of the former case 
for the same extension except for extension less than $\sim 1.0 \; {\rm \AA}$
(Fig. 12).

When the two strands of a base pair are stretched to some distance apart it
forces a part of the molecule into single stranded form and therefore creates
a boundary region separating the single and double stranded positions. The
base pairs of boundary region are open in the sense that the hydrogen bonds
are broken but the strands are not free to fluctuate because of neighbouring
bound base pairs. The peak in $F(y)$ corresponds to the energy needed to
create this boundary and its value is shown to depend on $k\rho$ and $D$. The
quantity $k\rho$ measures the barrier associated with the reduction in \d\
strand rigidity as one passes from ds\d\ to ssDNA. The value of $F(y)$ at the
peak for the potential parameters of set (i) is about three times larger than
that found for the set (ii); the reason being the large difference in the
value of $k\rho$ of the two sets of potential parameters. 

The measurements of the value of force at the peak in the $F(y)$ curve
of a ds\d\ molecule is, however, difficult as most experiments are carried
out under conditions where stretching of a base pair cannot be controlled
on the ${\rm\AA}$ scale. However, as we have shown in Figs. 8, 9 and 13,
the peak in the $F(y)$ curve can be made to occur at larger values of the
extension $y$ by replacing some of the base pairs by defect base pairs.
From these results it therefore seems possible to use defects to create the 
force barrier at such extensions of a base pair that the peak in $F(y)$ curve
can be measured directly in an experiment. 

The on-site potential on a defect site is represented by a potential that
has only a short-range repulsion and a flat part without well of the
Morse potential (Fig. 7). The coefficient $\rho$ of the anharmonic
term of the stacking interaction containing defect base pairs is also
suitably modified. With these modifications in the PB model and with the 
potential parameters of set (ii) we have calculated the $F(y)$ curves for 
different number of defect base pairs. The results shown in Figs. 8 \& 9 
are for the case in
which one of the ends base pair was stretched and in Fig. 13 for the 
case when the middle base pair was stretched. The qualitative features
of the curve $F(y)$ in the two cases are similar except in the case when
one base pair (i.e. the base pair which is being stretched) is a defect
base pair. This is because the barrier that gives peak in $F(y)$
curve in this case remains at the same location as in the defectless
case; the decrease in the value of force at the peak is primarily
due to loss of hydrogen bonds energy in the base pair.

It has been found (see Table 4) that in the presence of defects the peaks
in $F(y)$ shown in Fig. 13 is more than twice the corresponding peak
in Fig. 9. This has been attributed to difference in the entropic 
contributions which reduces the barrier height that gives the peak in 
$F(y)$ curve. 

Stretching a base pair at one of the ends of a ds\d\ molecule creates a
fork of Y shape which moves along the chain on increasing the
extension, $y$, of the end base pair. Its size for a given extension $y$
depends on the temperature as shown in Fig. 14. The number of base pairs
that get open on initiation of the formation of the fork depend rather
sensitively on the anharmonic term in the stacking energy. However,
after certain size of the fork the number of base pairs that get open
depend linearly on the extension $y$ as shown in Fig. 11.

When the middle (or any base pair far away from the ends) base pair
is stretched a bubble of ss\d\ with the shape of an eye is formed. In
a homogeneous ds\d\ molecule the bubble move symmetrically on both
sides on increasing the extension of the middle base pair. This may
not, however, happen in case of a heterogeneous ds\d\ molecule. The method
of matrix multiplications used in calculating the properties in the
constant extension ensemble can also be applied to a heterogeneous ds\d\
molecule.

\begin{center}
{\it Acknowledgement} 
\end{center}

The financial support from Council for Scientific and Industrial Research
(CSIR) and Department of Science and Technology (DST), Government of India,
New Delhi is acknowledged.

\end{document}